\documentclass[%
 reprint,
 amsmath,amssymb,
 aps,
floatfix,
]{revtex4-2}

\usepackage{graphicx}
\usepackage{dcolumn}
\usepackage{bm}
\usepackage{physics}
\usepackage{mathrsfs}

\graphicspath{ {./} }
\usepackage{subfigure}

\usepackage{filecontents}

\begin{document}

\preprint{APS/123-QED}

\title{Stability of bound states in multi-component DFT in absolute coordinate systems\\}

\author{Bander Linjawi}
\affiliation{Department of Mechanical Engineering,\\ King Abdullah University of Science and Technology, Thuwal 23955-6900, Saudi Arabia}




\date{\today}

\begin{abstract}
Homogeneous electron and nuclear gases are transformed to a localized trial density in absolute coordinates of the multi-component hamiltonian to determine the stability of forming bound states. Regions of stability were found both at the high density and low density regimes, where electron-nuclear correlations could play a critical role in the intermediate density regime. The use of Galilean coordinates is motivated for its use in density functional theory to develop kinetic and potential density functionals, from which suitable coordinate transformations to capture electron-nuclear correlations are applied.
\end{abstract}

\maketitle

\newcommand{\angstrom}{\mbox{\normalfont\AA}}
\section{\label{sec:level1}Introduction:\protect\\} 
Multi-component Density Functional Theory (MCDFT) allows for the description of electrons and nuclei on an equal footing quantum mechanically \cite{capitani1982non}. Such treatment is essential to capture electron-nuclear correlations, which play a role in various phenomena ranging from superconductivity to phase transitions and in warm dense matter \cite{lilia20222021,dornheim2025unraveling}. Various formulations have brought the theory to practice, such as the nuclear-electronic orbital approach at the Hartree-Fock \cite{pak2004electron} and DFT \cite{chakraborty2008development} level, and the time-dependent multi-component framework \cite{kreibich2008multicomponent}. Due to the absence of a natural external potential in multi-component theories, the translational invariance of the total hamiltonian induces constant electron and nuclear densities. A framework wherein the electronic coordinates are transformed to one fixed to the nuclear frame was developed \cite{kreibich2001multicomponent} using the often termed shape coordinates, and NEO-DFT necessarily introduces classical nuclei as external potentials to break the translational symmetry \cite{xu2022nuclear}. In the shape coordinates framework, it was further pointed that there is a freedom of choice for the nuclear coordinates \cite{butriy2007multicomponent} ranging from internal Jacobi coordinates to collective normal coordinates \cite{van2004first}. Moreover, the nuclear density is defined as the diagonal of the N-body density matrix without restriction to its marginal the single nuclei density, and the translational invariance is preserved for both intra-particle and inter-particle terms in the multi-component hamiltonian. The electronic subsystem however is described by its single electron density which is defined relative to the nuclear center of mass, retaining internal structure as in single component DFT, where the symmetry of the external potential fixes the frame. 

The N-body nature of the electronic subsystem can be reintroduced in this framework, which due to the larger inter-electronic overlap plays a more prominent role than that of the inter-nuclear interactions. This could also be done in the Galilean frame of the multi-component system without reliance on constant, marginal densities, however could deviate from the workhorse of single body DFT. The advantage of using Galilean coordinates is that electron-proton correlation and more generally exchange correlation functionals have been derived with respect to it, utilizing single body components. Moreover, coordinate transformations convert the intra-particle interactions to more than two-body, such in the Jacobi coordinates, or introduce coupling in the mass polarization term. This raises the difficulty in development of energy functionals in the multi-component DFT framework, which effects are shifted to the exchange-correlation of the intra- and inter-particle interactions \cite{kreibich2008multicomponent}. Additionally, higher rank functional theories such as Pair Density Functional Theory (PDFT) \cite{Ziesche1994,LevyZiesche2001} and theories employing geminals use distributions that non-trivially transform with the coordinates, making the Galilean coordinates easiest to work in. One could also utilize non-linear coordinate transformations, where the intra-particle interactions can be made block diagonal in the space of pair coordinates or their Jacobi counterparts. Nevertheless, finding an appropriate coordinate transformation that block diagonalizes both the kinetic and pair interactions is achieved, in many-body systems, only for model systems such as the harmonic oscillator. For Coulomb systems, the task is challenging, and the mass asymmetry in multi-component systems raises additional complications. 

To this end, this paper investigates the ability of a translationally invariant electron-nucelar hamiltonian (in Galilean coordinates) to admit bound states while absent from external potentials, and whether it admits a more stable ground state than a homogeneous system. We do so by studying a system of electronic and nuclear gases which give rise to a charge neutral system and convert them to localized density fluctuations. The variational parameter in our model is a decay length of the density fluctuations which we express in terms of the electron density, and when linked to the behavior of jellium we use as a measure of inhomogeneity, where lower densities denote higher inhomogeneity. We denote the system to favor a bound state if the localized trial densities lower the energy relative to the homogeneous state. It should be noted that this method utilizes a symmetry broken state similar to restricted or constrained treatments. Further, based on the single atom limit of our model, that is the hydrogen atom, we find that electron-nuclear correlations could play a prominent role in the many-body limit which our model can be improved to capture.

\section{\label{sec:level2}The model Hamiltonian and variational ansatz:\protect\\}
The electron-nuclear hamiltonian can be written in atomic units as
\begin{align}
H = \sum_{I}-\frac{\nabla_{I}^{2}}{2M} + \sum_{i}-\frac{\nabla_{i}^{2}}{2} + \frac{1}{2} \sum_{i<j} \frac{1}{|r_i-r_j|}\\\nonumber
 + \frac{1}{2} \sum_{I<J} \frac{Z^2}{|R_I-R_J|} - \sum_{I,i} \frac{Z}{|r_i-R_I|}
\end{align}

Under an arbitrary translation of the electron and ionic coordinates, the hamiltonian maintains. Further, the hamiltonian is written in the electronic and ionic Galilean or absolute lab frame coordinates, and the task to capture their correlation is moved to the form of the wave function. Using a plane wave product state for the electronic and ionic Fermi seas, the kinetic and intra-particle terms of the hamiltonian give the form of jellium for each other density, that is the nuclear background is treated explicitly, and for our purpose we retain the electron-nuclear potential. These terms can be written with respect to a single variable such as the electron density, and we further convert the electron-nuclear potential explicitly with respect to it. Namely, we equate the electronic and nuclear densities in this treatment, which allows the electron density to be the single descriptor. Evidently the Hartree terms of a system of electron and ion gases add up to zero in the product state of Fermi Seas, since locally the densities match with opposite charge, and only the inter-particle exchange survives to first order. Expanding about the free gas densities $\delta n(r) = n(r) - n_0,\, \delta m(r) = m(r) - m_0 $, such that $\int \!\mathrm dr\, \delta n(r) = 0$ and the same is the case for $\delta m(r)$, we ask whether the energy relative to the Fermi Seas lowers.

\begin{equation}
E[n_0+\delta n, m_0+\delta m] - E[n_0,m_0] < 0
\end{equation}

The kinetic energies require corresponding wave functions $\ket{\Psi}$ that evaluate to $n_0 + n(r)$ and $m_0 + m(r)$. We choose a Slater profile $\ket{\Psi}_{S}$ to represent the density fluctuations for convenience of evaluating the Coulomb integrals, and introduce the fluctuation with the electron-nuclear Fermi Seas $\ket{\Psi}_{FS}$ forming a product state of the two systems. 

\begin{equation}
\ket{\Psi} = \ket{\Psi}_{FS}\ket{\Psi}_{S}\\\nonumber
\end{equation}
\begin{align}
\ket{\Psi}_{S} = \Bigl(\frac{n_0m_0\alpha^{3}\beta^{3}}{(8\pi)^2}\Bigr)^{\!N/2}\prod_{i,I=1}^{N}e^{-\frac{(n_0^{\frac{1}{3}}\alpha r_i+m_0^{\frac{1}{3}}\beta R_I)}{2}}\\\nonumber
 =   \Bigl(\frac{n^2_0\alpha^{3}\beta^{3}}{(8\pi)^{2}}\Bigr)^{\!N/2}\prod_{i,I = 1}^{N}e^{-n_0^{\frac{1}{3}}\frac{(\alpha r_i+\beta R_I)}{2}}
\end{align}
\begin{align}
\delta n(r)  = \Bigl(\langle \sum_{i=1}^{N}\delta(r-r_i)\rangle_{S} - \langle \sum_{i=1}^{N}\delta(r-r_i)\rangle_{FS}\Bigl)\\\nonumber
 = n_0\Bigl(\frac{N\alpha^3}{8\pi}e^{-n_0^{\frac{1}{3}}\alpha r} - 1\Bigr)\\\nonumber
\delta m(r)  =  \Bigl(\langle \sum_{i=1}^{N}\delta(r-R_I)\rangle_{S} - \langle \sum_{i=1}^{N}\delta(r-R_I)\rangle_{FS}\Bigl)\\\nonumber
 = n_0\Bigl(\frac{N\beta^3}{8\pi}e^{-n_0^{\frac{1}{3}}\beta R} - 1\Bigr)
\end{align}

We set $m_0 = n_0$ by setting their charges equal $Z=1$ to maintain charge neutrality, otherwise the nuclear Sea and localized state would have to be contracted by $Z^{\frac{1}{3}}$. It should be noted that the localized wave function and the densities it induces break the full translational symmetry, and while the hamiltonian commutes with such symmetry, its ground state spectrum may reside in a symmetry broken state. Further, the density fluctuation studied is at the extreme limit of converting the entire gases into localized states, retaining the product state nature in a many-body setting. Had we relaxed this, interactions between the gases and the fluctuation would have to be accounted for. In essence, we are asking whether an uncorrelated electron-nuclear system favors the gas or the broken symmetry state, and this treatment should naturally extend to accounting for correlations. Furthermore, we partition the energy explicitly in terms of Hartree and exchange terms, and as mentioned neglect correlations in the mean time in our product state analysis

\begin{equation}
E = \langle T_r \rangle + \langle T_R \rangle + V_H[n] + V_H[m] + V_H[n,m] + V_{x}[n] + V_{x}[m]
\end{equation}

Further, since we are describing a product state, we approximate the exchange by the Local Density Approximation (LDA) for the intra-particle exchange, given by

\begin{equation}
V_{x}[\rho] = -\frac{3}{4} \Bigl(\frac{3}{\pi}\Bigr)^{\frac{1}{3}} \int \!\mathrm d^3r\, \rho^{\frac{4}{3}}(r)
\end{equation}

The exchange in local form does not change the aim of this study, as the main point of investigation is the applicability of the translationally invariant hamiltonian with respect to inertial coordinates in allowing bound states, and the local LDA exchange is written with respect to such coordinates. However, while the LDA exchange is a large term and stabilizes the nuclear gas, at large $N$ it may over stabilize the bound state. The Hartree terms are also evaluated in lab frame coordinates. Further, it is easy to show that the difference in energy between the two states can now be expressed in the following where the Hartree energy is quadratic in the localized densities by construction \footnote{$E[n_0+\delta n, m_0+\delta m] = \langle H_{FS}\otimes I_{S} + I_{FS}\otimes H_{S}\rangle - \langle H_{F}\otimes I_{S}\rangle = \langle I_{FS}\otimes H_{S}\rangle$}. We express the total energy difference as

\begin{flalign}
E[n_0+\delta n, m_0+\delta m] - E[n_0,m_0]  = \\\nonumber
\langle T_r + T_R \rangle_{SD}- \langle T_r + T_R \rangle_{FS}\\\nonumber
+\frac{1}{2} \int\int \!\mathrm d^3r\, \!\mathrm d^3r'\, \frac{n(r)n(r')}{|r-r'|}\\\nonumber
 \\\nonumber
  + \frac{1}{2} \int\int \!\mathrm d^3R\, \!\mathrm d^3R'\, \frac{m(R)m(R')}{|R-R'|}\\\nonumber 
\\\nonumber
 - \int\int \!\mathrm d^3r\, \!\mathrm d^3R\, \frac{n(r)m(R)}{|r-R|} - \frac{3}{4} \Bigl(\frac{3}{\pi}\Bigr)^{\frac{1}{3}} \int \!\mathrm d^3r\, n^{\frac{4}{3}}(r) \\\nonumber
\\\nonumber
- \frac{3}{4} \Bigl(\frac{3}{\pi}\Bigr)^{\frac{1}{3}} \int \!\mathrm d^3R\, m^{\frac{4}{3}}(R)+ \frac{3}{4} \Bigl(\frac{3}{\pi}\Bigr)^{\frac{1}{3}}\Bigl(n_0^{\frac{1}{3}} + m_0^{\frac{1}{3}}\Bigr)N
\end{flalign}

where the Hartree terms cancel exactly at the homogeneous density state. The Hartree term can readily be evaluated by classical treatments \cite{griffiths2018introduction} which is generally given by

\begin{align}
\int\int \!\mathrm d^3r\, \!\mathrm d^3r'\, \frac{e^{-\alpha r}e^{-\alpha r'}}{|r-r'|} = \frac{20\pi^2}{\alpha^5}\\\nonumber
\int\int \!\mathrm d^3r\, \!\mathrm d^3r'\, \frac{e^{-\alpha r}e^{-\beta r'}}{|r-r'|} = \frac{32\pi^{2}(\alpha^2+3\alpha\beta+\beta^2)}{\alpha^2\beta^2(\alpha+\beta)^{3}}
\end{align}

\begin{align}
V_H[n]  =  \frac{5n_0^{\frac{1}{3}}\alpha}{32}N^2;\,V_H[m] =   \frac{5n_0^{\frac{1}{3}}\beta}{32}N^2\\\nonumber
V_H[n,m] = -\frac{n_0^{\frac{1}{3}}\alpha\beta(\alpha^2+3\alpha\beta+\beta^2)}{2(\alpha+\beta)^3}N^2 
\end{align}

The exchange terms can be readily evaluated

\begin{align}
V_X[n] = - \frac{3}{4} \Bigl(\frac{3}{\pi}\Bigr)^{\frac{1}{3}} \int \!\mathrm d^3r\, n^{\frac{4}{3}}(r) = - \Bigl(\frac{3}{4}\Bigr)^4\Bigl(\frac{3}{8\pi^2}\Bigr)^{\frac{1}{3}}n_0^{\frac{1}{3}}\alpha N^{\frac{4}{3}}
\end{align}
\begin{align}
V_X[m] = -\Bigl(\frac{3}{4}\Bigr)^4\Bigl(\frac{3}{8\pi^2}\Bigr)^{\frac{1}{3}} n_0^{\frac{1}{3}}\alpha N^{\frac{4}{3}}
\end{align}

and the kinetic energies of the localized states are

\begin{align}
\langle T_r + T_R \rangle_{S} =  \frac{n_0^{\frac{2}{3}}}{8}\Bigl(\alpha^2 +\frac{\beta^{2}}{M}\Bigr) N\\\nonumber
\langle T_r + T_R \rangle_{FS} =   \frac{3}{10}\Bigl(3\pi^2\Bigr)^{\frac{2}{3}}n_0^{\frac{2}{3}}\Bigl(1 + \frac{1}{M}\Bigr)N
\end{align}

The explicit scaling with $N$ is denoted for each term, which surprisingly differs for the localized state than for the homogeneous state. Namely, the localized state exchange scales stronger by $N^{\frac{1}{3}}$. A question to pose is whether the exchange functional should be size consistent, that is

\begin{equation}
V_X[n_1 + n_2] = V_X[n_1] + V_X[n_2]
\end{equation}

The exchange arises from a two-body interaction, therefore we expect the above equality not to hold as interactions between subsystems do occur. However, in our explicit choice of a product state, the kinetic energy is non-interacting as they are chosen to be eigenstates of it, which should yield the same scaling as the Thomas Fermi functional ($O(N)$). This is not the case, rather we obtain a factor of $N^{\frac{2}{3}}$ too large when the Thomas Fermi functional is used compared to the expected linear N scaling obtained by the above expectation value. Interestingly, resolving the density by orbitals of the localized state obtains the expected scaling when using density functionals, while it reduces the scaling by $N^{\frac{1}{3}}$ and $N^{\frac{2}{3}}$ for the exchange and kinetic energy of the homogeneous state respectively. It is as if the homogeneous state abides by the corresponding density functionals built with respect to while the localized state favors to be orbital resolved. In other words, the density functionals add correlations to the localized state which the product state did not warrant ($T_{TF}[n_1 + n_2] \neq T_{TF}[n_1] + T_{TF}[n_2]$). The Hartree term at large $N$ is indifferent to the orbital or density treatments as at the orbital level scales as $N(N-1)/2$. $N$-dependent normalizations of the density functionals are found to be beneficial \cite{clay2025approximate} and scaling behavior of density functionals has been extensively studied \cite{fabiano2013relevance}. In this study, we fix the scaling to give the same scaling as the homogeneous state. The energy difference becomes

\begin{align}
\Delta V_{H} = -n_0^{\frac{1}{3}}\Bigl(\frac{\alpha\beta(\alpha^2+3\alpha\beta+\beta^2)}{2(\alpha+\beta)^3}- \frac{5}{32}(\alpha + \beta)\Bigr)N^{2}\\\nonumber
= -\frac{0.6204}{r_s}\Bigl(\frac{\alpha\beta(\alpha^2+3\alpha\beta+\beta^2)}{2(\alpha+\beta)^3}- \frac{5}{32}(\alpha + \beta)\Bigr)N^{2}
\end{align}

\begin{align}
\Delta{V_X} = \frac{3}{2}\Bigl(\frac{3}{\pi}\Bigr)^{\frac{1}{3}}n_0^{\frac{1}{3}}\Bigl(1-\Bigl(\frac{3}{4}\Bigr)^{3}\Bigl(\frac{3}{8\pi^{2}}\Bigr)^{\frac{1}{3}}\alpha\Bigr)N\\\nonumber
 = \frac{0.916}{r_s}\Bigl(1 - 0.071(\alpha + \beta)\Bigr)N
\end{align}

\begin{align}
\Delta{T} = \Bigl(\frac{n_0^{\frac{2}{3}}}{8}\Bigl(\alpha^2 + \frac{\beta^2}{M}\Bigr)  - \frac{3}{10}\Bigl(3\pi^2\Bigr)^{\frac{2}{3}}n_0^{\frac{2}{3}}\Bigr)N\\\nonumber
 = \Bigl(\frac{0.0481}{r^2_s}\Bigl(\alpha^2 + \frac{\beta^2}{M}\Bigr) - \frac{1.105}{r^2_s}\Bigr)N
\end{align}

where we used the relation $n_0 = \frac{3}{4\pi r^{3}_s}$ and the Winger-seitz radius $r_s$ is a measure of inhomogeneity since the kinetic and potential terms scale differently with respect to it. The value of $\alpha$ and $\beta$ can be found in terms of $r_s$ by minimizing the energy difference with respect to the exponents with increasing $r_s$

\begin{align}
\Delta E(\alpha,\beta) = \Bigl(\frac{0.0481}{r^2_s}\Bigl(\alpha^2 + \frac{\beta^2}{M}\Bigr) - \frac{1.105}{r^2_s}\Bigl(1 + \frac{1}{M}\Bigr)\Bigr)N \\\nonumber
  + \frac{0.916}{r_s}\Bigl(1 - 0.071(\alpha+\beta)\Bigr)N \\\nonumber
-\frac{0.6204}{r_s}\Bigl(\frac{\alpha\beta(\alpha^2+3\alpha\beta+\beta^2)}{2(\alpha+\beta)^3}- \frac{5}{32}(\alpha + \beta)\Bigr)N^{2}
\end{align}

\begin{figure}[htbp]
\centering
\includegraphics[width=\columnwidth]{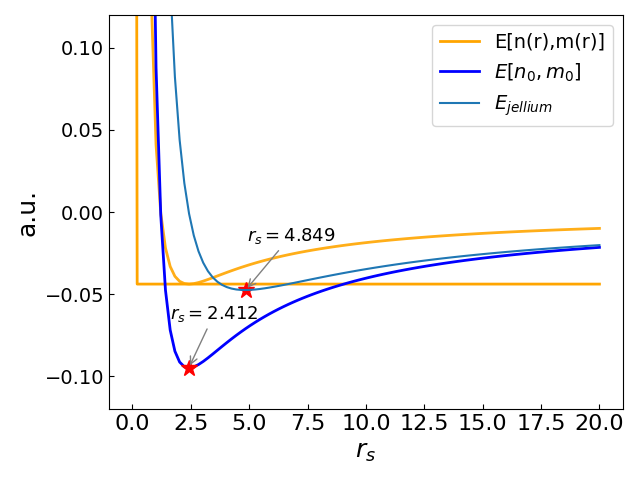}
\caption{Energies of the bound state and the homogeneous state with $r_s$, showing the regions of stability of the bound state both at high density and low density regimes and the symmetrical behavior of the variational ansatz. The electronic jellium energy is shown for comparison.}
\label{fig:dE}
\end{figure}

The energy difference was minimized using the BFGS optimization algorithm, and $N$ was set to $1000$ to approach the thermodynamic limit. We noticed larger values to induce considerable fluctuations approaching large $r_s$, therefore we restricted to below $r_s=20$ in this study. The optimal $r_s$ for a gas of electrons and nuclei at equal densities was found to be $r_s=2.412$ which corresponds to the minima. Had the nuclear exchange was set to 0, we obtain an $r_s=4.812$ as in the electron jellium case. Although we have introduced different decay lengths for the electrons and nuclei, the optimal model proved to be $r_s$ independent, with the energy of the localized state found to be $-0.0439$ Ha. At an equal average density of the nuclear and electronic components, the model variationally gives the same local density profiles $\alpha=\beta$. Fig. \ref{fig:dE} shows the energies of the localized and homogeneous gas states with $r_s$ per particle. The flat orange line is the variational solution, while the orange curve represents a variational solution where $\alpha$ the electronic width was fixed with respect to the free gas minimum at $r_s=2.412$ while $\beta$ was optimized. This was done at $\alpha^{*}=\beta^{*}=1.352\,r_s$. The curve minima traces the flat bound with increasing $\alpha$ at different values of $r_s$. This behavior is likely arising from the symmetry of the Slater wave function profiles of the electronic and nuclear subsystems used, causing the preference of perfect cancellation of the Hartree terms. Another plausible effect impacting the observed symmetrical behavior is that both densities scale with $r_s$ identically, although $\alpha$ and $\beta$ were varied. To investigate this, we removed the $n_0=m_0$ condition and scaled the nuclear densities by $m_0$ with a different Wigner Seitz radius $r_{s2}$ while keeping the particle number fixed. The $r_s$ independence did not break, and therefore we attribute it to the matching shapes of the electronic and nuclear orbitals. Furthermore, surprisingly, the bound state stabilizes in the high density regime, where the free gas kinetic energy amplifies. At the critical value of $r_s =2.412$, the energy difference is found to be $0.051$ Ha, almost equal to the energy of the localized state itself. Therefore, another study could perturb the system from this starting point. At this minima, we find $\Delta V_X = 0.102$ Ha while $\Delta T = -0.051$ Ha. 

\section{\label{sec:level3}Inclusion of the correlation energy:\protect\\}
Different from the relative coordinates hydrogen orbitals, the product state of Slater determinants neglects the correlation energy of a hydrogen atom, or the difference between a product state and the true relative coordinates state. Particularly, we obtain for the electron-nuclear Hartree term at the optimal $\alpha^{*}$ a value of $-0.9108$ Ha, giving a correlation energy of $-0.0892$ Ha or $-0.0446$ Ha per particle, which is comparable to the required energy to stabilize the bound state at the free gas minimum. At the many-body level, the Hartree-Fock wave functions also miss correlations. In order to capture them, the pair density becomes central, and is defined as

\begin{equation}
P_{2}(r_1,r_2) = N(N-1) \int\!\mathrm{d}r_3...\mathrm{d}r_N |\Psi|^{2}
\end{equation}

which is often coordinate transformed from the Galilean coordinates, and likely is not translationally invariant (in fact as we show below it can depend on the center of mass). Nevertheless, it remains two-body, of the same dimension as the Coulomb interaction. This detail has been utilized to develop electronic correlation functionals \cite{colle1975approximate} and electron-nuclear correlation functionals \cite{yang2017development}. In the development of electron-nuclear correlation functionals, the starting point is a product of electron and nuclear Hartree-Fock wave functions multiplied by a Jastrow function that introduces correlations between the two components. 

\begin{equation}
\ket{\Psi} = \ket{\Psi}_{HF}\prod_{i,I}(1-\phi(r_{i},R_{I}))
\end{equation}


The Jastrow function is naturally written in the center of mass and relative positions coordinates

\begin{equation}
\phi(r_{rel},R_{CM}) = e^{-\gamma^2r_{rel}}\Bigl(1-\Phi(R_{CM})\bigl(1 + \frac{r_{rel}}{2}\bigr)\Bigr)
\end{equation}

with a coordinate transformation

\begin{align}
\begin{pmatrix} R_{CM} \\ r_{rel} \end{pmatrix} = \begin{pmatrix} \frac{M}{M+m} &  \frac{m}{M+m} \\ 1 &  -1 \end{pmatrix}  \begin{pmatrix} R \\ r \end{pmatrix}
\end{align}

with a Jacobian determinant of 1 denoted generally as $|J|$. The coordinate transformation of the wave functions and the pair densities is such that the distributions remain normal, with the transformation of $B^2 = |J| A^2$ where $A$ and $B$ are orbital normalization coefficients in the original and transformed coordinate systems respectively. Linear coordinate transformations can be thought of as anisotropically scaling each coordinate independently, or Affine transformations, and the distribution functions are homogeneous of degree $k$ depending on their shape $f(r) = \lambda^{k}f(\lambda r)$. Importantly, the Coulomb energy scales with the pair density as

\begin{equation}
V[P_2(r_1,r_2)] = \lambda V[P_2(\lambda r_1, \lambda r_2)]
\end{equation}

The important insight that eases the evaluation of the correlation as well as the general Coulomb interaction is that the coordinate transformations are performed pair by pair, allowing to preserve the two-body Coulomb kernel structure. This is allowed by two factors, for the general Coulomb interaction it strictly depends only on the pair-density which is a two-body quantity, and for the correlation energy the Jastrow function can be approximated as such as well \cite{colle1975approximate}. This circumvents the need for many-body coordinate transformations where the interactions may become non-trivial. To investigate this at the many-body level, we incorporate a DFT based correlation functional for our product state and utilize the epc \cite{tao2019multicomponent} functional without gradient terms since they evaluate to 0 for Fermi Seas

\begin{equation}
V_C[n, m] = - \int\!\mathrm d^3s\, \frac{n(s)m(s)}{a-b\bigl[n^{\frac{1}{2}}(s)m^{\frac{1}{2}}(s)\bigr] + c\bigl[n(s)m(s)\bigr]}
\end{equation}

where $a,b,c$ are taken from Ref. \cite{yang2017development} as  2.35, 2.4, and 3.2 respectively. The coordinate $s$ is the center of mass coordinate $s = R'_{CM} =\frac{1}{2}(r + R)$ and the functionals were developed following the Colle-Salvetti treatment of electron correlations \cite{colle1975approximate}. Importantly, a many-body reduction to two-body coordinates was achieved by utilizing the mean-value theorem for the electron coordinates, of which the two-body Coulomb integrals can be converted further to single body density functionals. In the original formulation of the Colle-Salvetti correlation energy, there was no complication arising from mass differences as it treated the electronic system. The electron-proton correlation functional derived from Hartree-Fock determinants does reduce to the Colle-Salvetti form only after appropriately mass-weighting the lab frame coordinates by the following coordinate transformation (we retain the electronic mass $m$ for ease of read)

\begin{equation}
\begin{pmatrix} R' \\ r' \end{pmatrix} = \begin{pmatrix} \frac{M+m}{2M} & 0 \\ 0 &  \frac{M+m}{2m} \end{pmatrix} \begin{pmatrix} R \\ r \end{pmatrix}
\end{equation}

which makes the conversion from Galilean frame to center of mass and relative coordinates frame to be

\begin{align}
\begin{pmatrix} R'_{CM} \\ r'_{rel} \end{pmatrix} = \begin{pmatrix} \frac{M}{M+m} &  \frac{m}{M+m} \\ 1 &  -1 \end{pmatrix} \begin{pmatrix}  \frac{M+m}{2M} & 0 \\ 0 &  \frac{M+m}{2m} \end{pmatrix}  \begin{pmatrix} R \\ r \end{pmatrix}\\\nonumber
\begin{pmatrix} R'_{CM} \\ r'_{rel} \end{pmatrix} = \begin{pmatrix}\frac{1}{2} &  \frac{1}{2} \\  \frac{M+m}{2M}  &  -\frac{M+m}{2m} \end{pmatrix}\begin{pmatrix}  R \\ r \end{pmatrix}
\end{align}

with the inverse Jacobi determinant of transformation of $4\frac{\mu}{M+m}$, additionally our coordinate system has an additional $\alpha^{2}$ to bring the coordinates to $R'_{CM}$, but this Jacobi is not suitable for the chosen parameters. Moreover, the product of the densities arising from the product of Slater profiles $n(r)m(R)$ becomes 

\begin{equation}
n(r)m(R) \propto e^{-\alpha(r+R)} = e^{-2\alpha R'_{CM}} = e^{-2\alpha s}
\end{equation}

We evaluate the correlation energies of the density fluctuations at the ground state of $r_s = 2.412$. We obtain $-0.012$ Ha for the localized state while for the gas state $-0.0462$ Ha, indicating that the gas state remains more stable. It is possible that the gas state correlation is overestimated in this treatment.

\section{\label{sec:level4}Conclusion:\protect\\}
An electron and nuclear density localized state is assessed against a homogeneous electron and nuclear gas ground state to determine the stability of a multi-component bound state. At an equal average density, the Slater profiles variationally matched, and breaking this constraint by fixing the width of one component gave an unstable bound state in the density regime studied, and incorporation of electron-nuclear correlations will likely play a crucial role in promoting stability. Galilean or lab frame coordinates are motivated for ease of construction and application of exchange-correlation single density functionals, as well as in treatments of higher rank density functional theories upon which electron-nuclear correlation functionals are derived from. Further, the employed model determined a stability window of bound states below $r_s = 1.4$ and above $r_s = 10$ in the regime investigated, and can be improved by using a better starting ansatz. The two states are currently only heuristically connected by maintaining the same density globally and thus particle number, and a more realistic treatment perturbs one state towards the other as well as describing the electrons and nuclei with different shapes. Further investigations are underway.
\bibliography{main}
\bibliographystyle{unsrt}

\end{document}